\documentclass[a4paper]{jpconf}
\usepackage{graphicx}

\usepackage{lineno}

\begin{document}
\title{Strangeness production in \mbox{p--Pb} and \mbox{Pb--Pb} collisions with ALICE at LHC}

\author{Domenico Colella$^{1,2}$ for the ALICE Collaboration}

\address{$^{1}$ European Organization for Nuclear Research (CERN)}
\address{$^{2}$ Institute of Experimental Physics, Slovak Academy of Sciences (IEP SAS)}

\ead{domenico.colella@cern.ch}

\begin{abstract}
The main goal of the ALICE experiment is to study the properties of the hot and dense medium created in ultra-relativistic
heavy-ion collisions. The measurement of the \mbox{(multi-)strange} particles is an important tool to understand particle 
production mechanisms and the dynamics of the quark-gluon plasma (QGP). 
We report on the production of K$^{0}_{S}$, \mbox{$\Lambda$($\overline{\Lambda}$)}, \mbox{$\Xi^{-}$($\overline{\Xi}^{+}$)} 
and \mbox{$\Omega^{-}$($\overline{\Omega}^{+}$)} in proton-lead (\mbox{p--Pb}) collisions at $\sqrt{s_{\rm NN}}$ 
= 5.02 TeV and lead-lead (\mbox{Pb--Pb}) collisions at $\sqrt{s_{\rm NN}}$ = 2.76 TeV measured by ALICE at the LHC.
The comparison of the hyperon-to-pion ratios in the two colliding systems may provide insight into strangeness production 
mechanisms, while the comparison of the nuclear modification factors helps to determine the contribution of 
initial state effects and the suppression from strange quark energy loss in nuclear matter. 
\end{abstract}

\section{Introduction}
Strangeness production has been extensively studied in relativistic heavy-ion collisions. Its measurement represents
an important tool to investigate the properties of the strongly interacting system created in the collision,
as there is no net strangeness content in the initially colliding nuclei. 
In these proceedings we briefly discuss the technique used to identify K$^{0}_{S}$, \mbox{$\Lambda$($\overline{\Lambda}$)}, 
\mbox{$\Xi^{-}$($\overline{\Xi}^{+}$)} and \mbox{$\Omega^{-}$($\overline{\Omega}^{+}$)} and to measure their transverse 
momentum ($p_{\rm T}$) spectra. 
Then we describe the main results about baryon anomaly, strangeness nuclear modification factor and strangeness enhancement.

\section{Strange particle identification with the ALICE detector}
The ALICE detector was designed to study heavy-ion physics at the LHC. 
At mid-rapidity, tracking and vertexing are performed using the Inner Tracking System (ITS), consisting of six layers of 
silicon detectors, and the Time Projection Chamber (TPC). The two innermost layers of the ITS and the V0 detector (scintillation 
hodoscopes covering the forward pseudo-rapidity region on either side of the interaction point) are used for triggering. The V0 
is also used to estimate centrality in \mbox{Pb-–Pb} collisions as well as multiplicity in \mbox{p–-Pb} collisions. A complete 
description of the ALICE sub-detectors can be found in \cite{JINST}.

The single-strange (K$^{0}_{S}$ and $\Lambda$) and multi-strange ($\Xi$ and $\Omega$) particles
are reconstructed via their characteristic weak decay topologies into two and three particles, respectively. Tracks reconstructed by 
the TPC and the ITS are combined to select candidates satisfying a set of geometrical criteria. In addition, particle 
identification is performed by a selection on the specific energy loss in the TPC for the daughter tracks.
Particle yields as a function of $p_{\rm T}$ are determined, in various multiplicity/centrality intervals, using an invariant 
mass analysis.
Acceptance and efficiency corrections are calculated using Monte Carlo simulations. For more details, we refer to 
\cite{multsbPbPb,lfpPb,multsbpPb}.

\section{Results}
Transverse momentum spectra for K$^{0}_{S}$, \mbox{$\Lambda$($\overline{\Lambda}$)}, \mbox{$\Xi^{-}$($\overline{\Xi}^{+}$)} and 
\mbox{$\Omega^{-}$($\overline{\Omega}^{+}$)} have been published in \cite{singlsPbPb,multsbPbPb} for \mbox{Pb--Pb} collisions and 
in \cite{lfpPb,multsbpPb} for \mbox{p--Pb} collisions. In both systems the hardening of the spectral shape from low- to 
high-multiplicity collisions has been observed. 
Moreover, in the context of Blast Wave model, it has been shown that spectra for K$^{0}_{S}$ and $\Lambda$ are well predicted 
using parameters from a simultaneous fit to $\pi^{\pm}$, K$^{\pm}$ and p spectra in high multiplicity \mbox{Pb--Pb} and \mbox{p--Pb} 
collisions. This is also true for $\Xi$ and $\Omega$ in \mbox{p--Pb} collisions, while in \mbox{Pb--Pb} collisions
they cannot be described in a common freeze-out scenario, as they would require a lower mean transverse flow velocity and a 
higher kinetic freezeout temperature to be described properly.

\begin{figure}[t]
\centering
\includegraphics[width=20pc,clip]{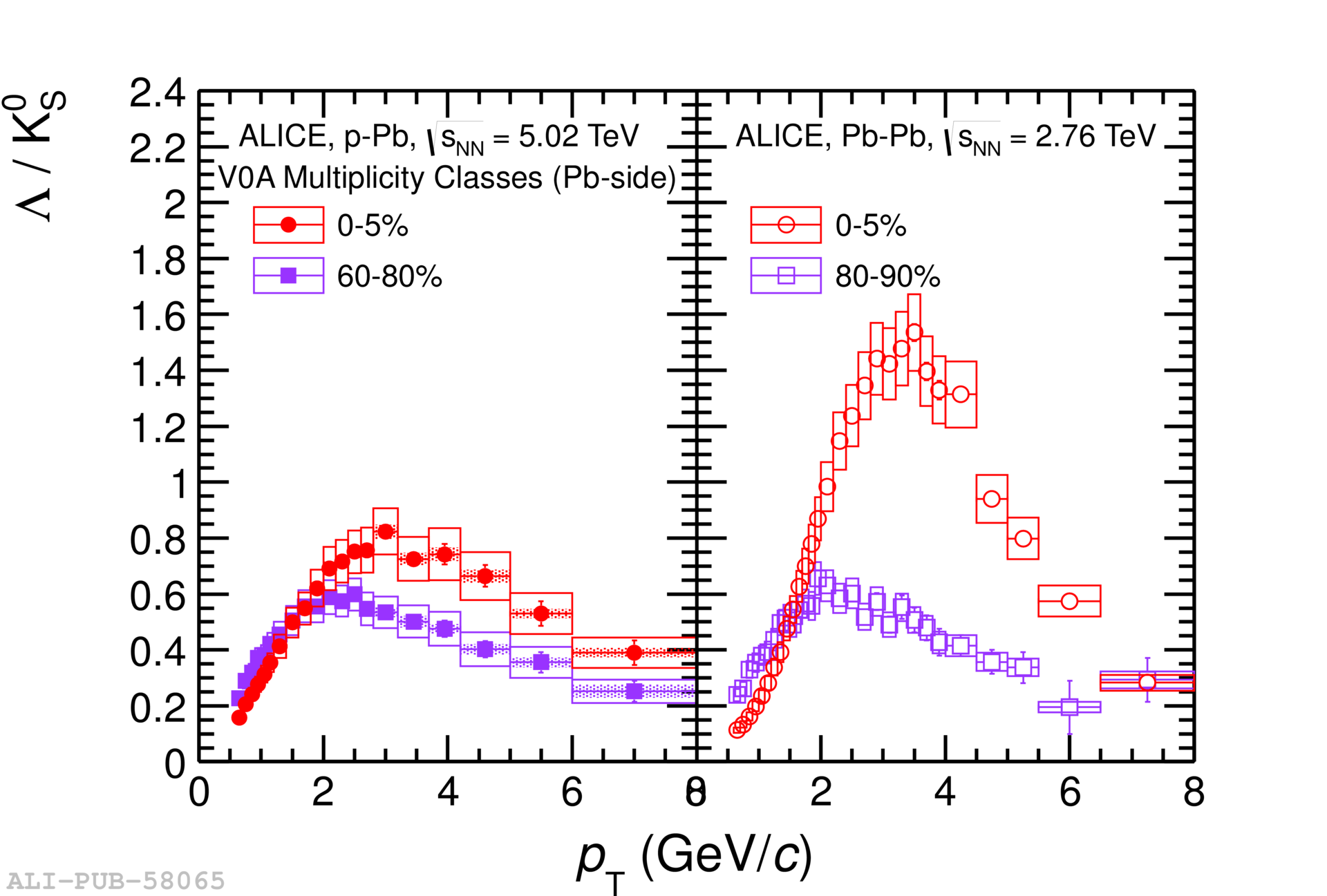}\hspace{1pc}%
\begin{minipage}[b]{12pc}\caption{($\Lambda$+$\overline{\Lambda}$)/K$^{0}_{S}$ as a function of $p_{\rm T}$ in 
                                   the highest and a low-multiplicity class in \mbox{p--Pb} (left) and \mbox{Pb--Pb} 
                                   (right) collisions. Open boxes are total systematic uncertainties and shaded boxes 
                                   represent the uncorrelated-over-multiplicity component.}
\label{fig-0}
\end{minipage}
\end{figure}

The $p_{\rm T}$-differential ($\Lambda$+$\overline{\Lambda}$)/K$^{0}_{S}$ ratio is shown in Fig. \ref{fig-0} in the highest
and a low-multiplicity class for \mbox{p--Pb} and \mbox{Pb--Pb} collisions. The  shape and the multiplicity dependence are 
qualitatively similar in the two collision systems. In \mbox{Pb--Pb} collisions this phenomenon has been interpreted
as a redistribution of baryons and mesons in momentum, when centrality increases, a consequence of the increased
radial flow \cite{singlsPbPb}. The flattening of the ($\overline{p}$+$p$)/$\phi$ ratio as a function of $p_{\rm T}$ in the 0-10\% 
central \mbox{Pb--Pb} collisions is further evidence supporting an hydrodynamical evolution of the system. In addition this data
can also be described by parton recombination models \cite{resPbPb}.

\begin{figure}[t]
\centering
\includegraphics[width=14pc,clip]{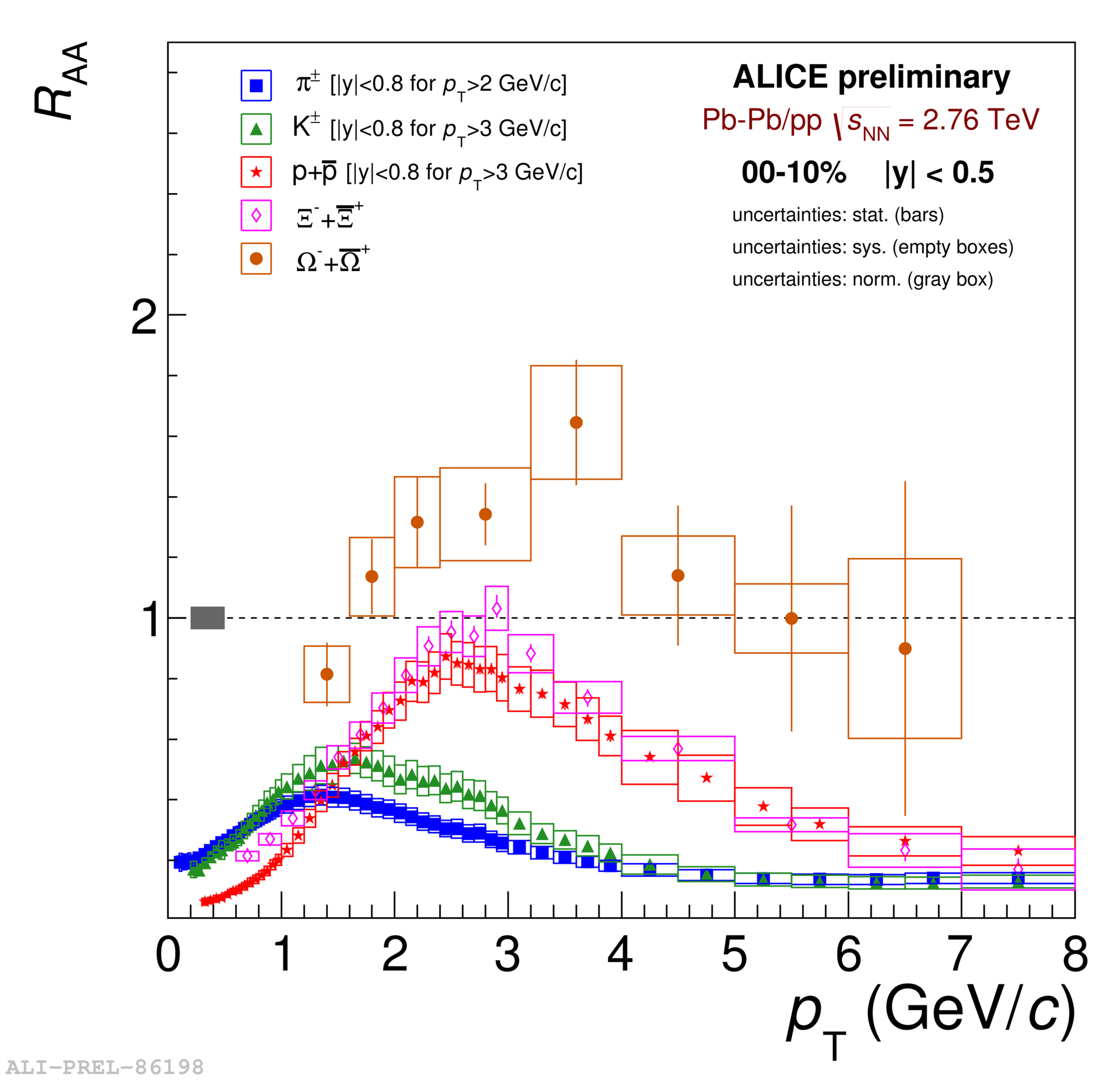}
\includegraphics[width=19pc,clip]{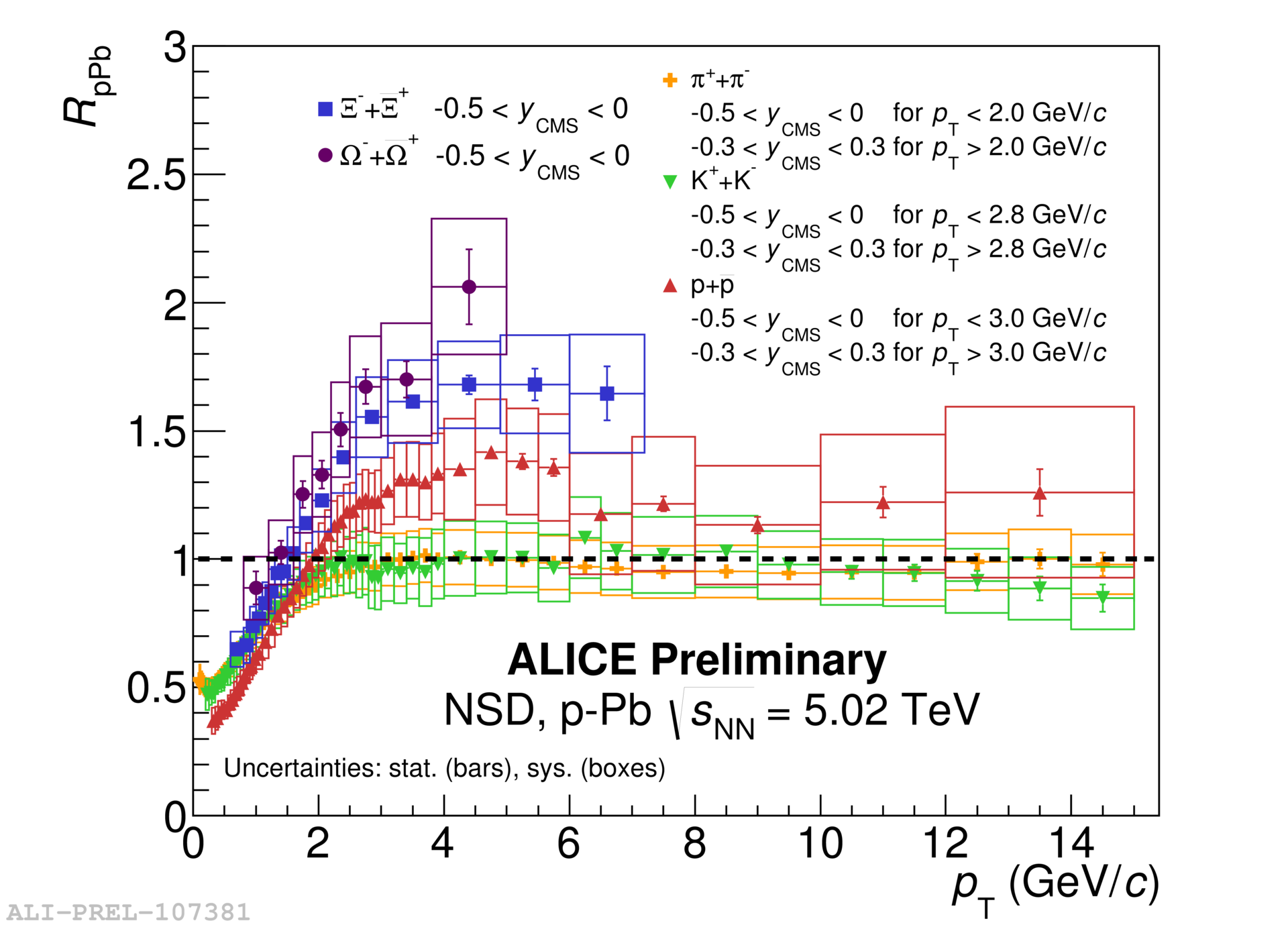}
\caption{$R_{\rm pPb}$ in \mbox{p--Pb} collisions (left) and $R_{\rm AA}$ in 0-10\% most central 
         \mbox{Pb--Pb} collisions (right) as a function of $p_{\rm T}$ for $\pi$, K, p, $\Xi$ and $\Omega$ particles. Open 
         boxes are total systematic uncertainties.}
\label{fig-2} 
\end{figure}

The nuclear modification factor is defined as the ratio of the $p_{\rm T}$ spectra in \mbox{Pb--Pb} ($R_{\rm AA}$) 
or \mbox{p--Pb} ($R_{\rm pPb}$) and in pp collisions scaled by the number of nucleon-nucleon collisions. It has been 
shown that the strong suppression of hadron production at high $p_{\rm T}$ observed at the LHC in \mbox{Pb--Pb} 
collisions is not due to an initial-state effect but may be due to jet quenching in hot QCD matter \cite{unidpPb}.
In Fig. \ref{fig-2} the $R_{\rm AA}$ as a function of $p_{\rm T}$ for multi-strange baryons in central 
(0-10\%) collisions is shown and compared with those for lighter hadrons ($\pi$, K and p). $R_{\rm AA}$ for 
$\Xi$ is similar to the one for p, especially at high $p_{\rm T}$ \mbox{($>$ 6 GeV/{\it c})}, while $\pi$ and K at low $p_{\rm T}$
\mbox{($<$ 3 GeV/{\it c})} follow a clearly different trend, which can be interpreted as a consequence of having different masses 
in a radial flow scenario.
$R_{\rm AA}$ for $\Omega$ is above unity, which might be the result of the larger strangeness enhancement compared to 
$\Xi$ \cite{multsbPbPb}. In the right canvas of the same figure, the $R_{\rm pPb}$ as a function of $p_{\rm T}$ in minimum-bias 
\mbox{p--Pb} collisions is shown for the same particles. In this case, there is no suppression for $\pi$, K and p at high 
$p_{\rm T}$ ($>$ 6 GeV/{\it c}), as verified for unidentified particles \cite{unidpPb}.
In the so-called Cronin region (2 $<$ $p_{\rm T}$ $<$ 6 GeV/{\it c}) an increase of the $R_{\rm pPb}$ with 
the mass and the strangeness content of the particles is visible.

\begin{figure}[t]
\centering
\includegraphics[width=16pc,clip]{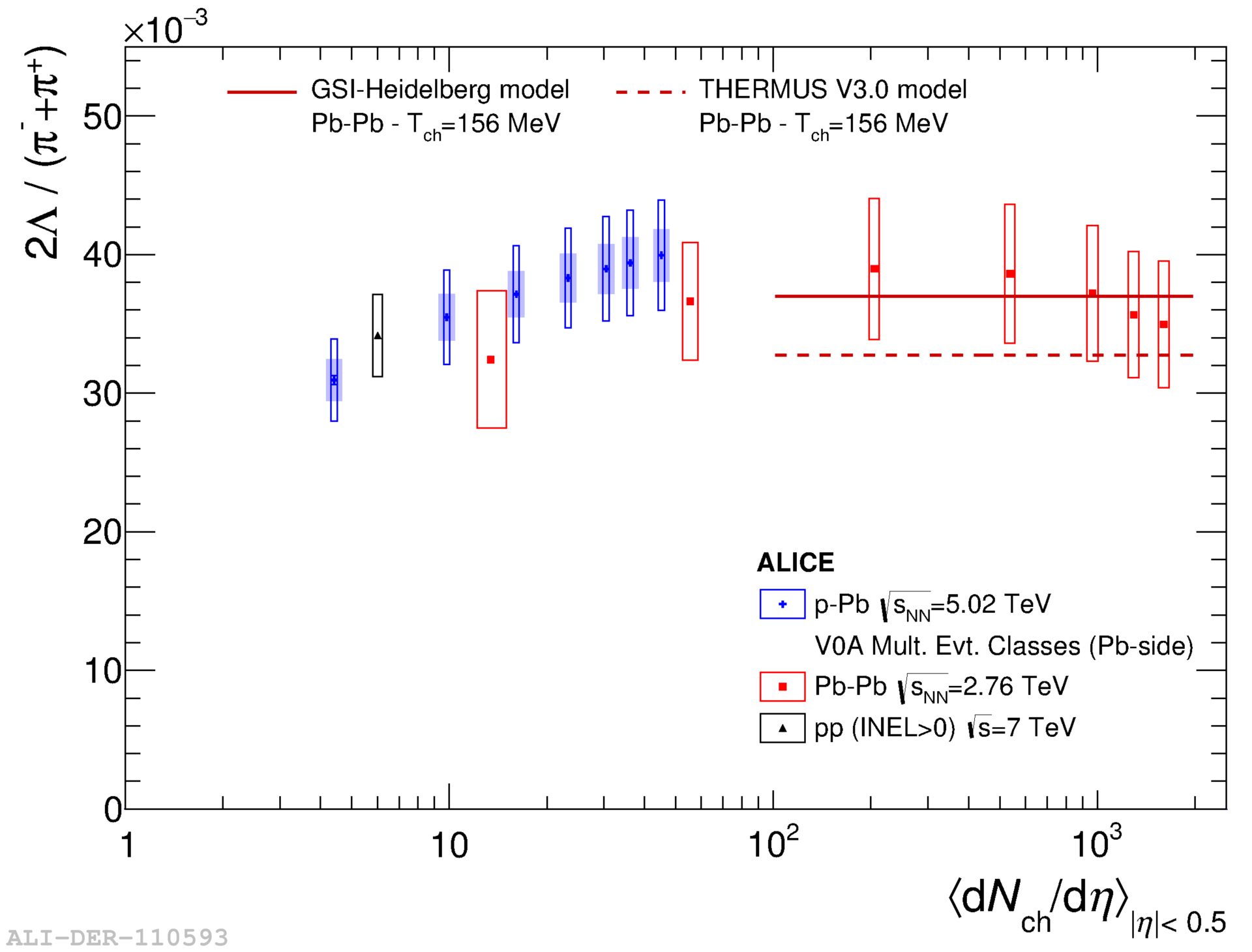}
\includegraphics[width=16pc,clip]{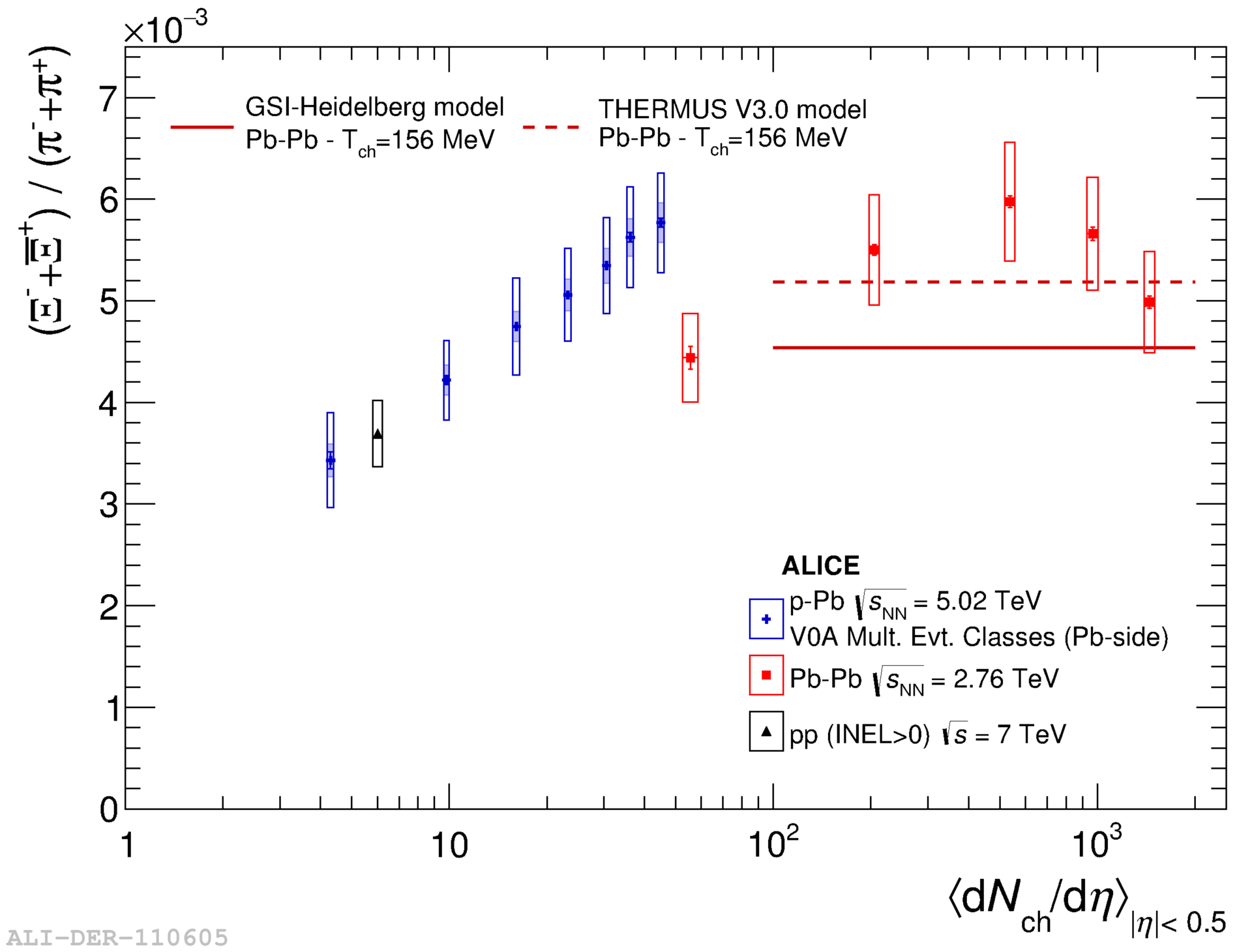}
\includegraphics[width=16pc,clip]{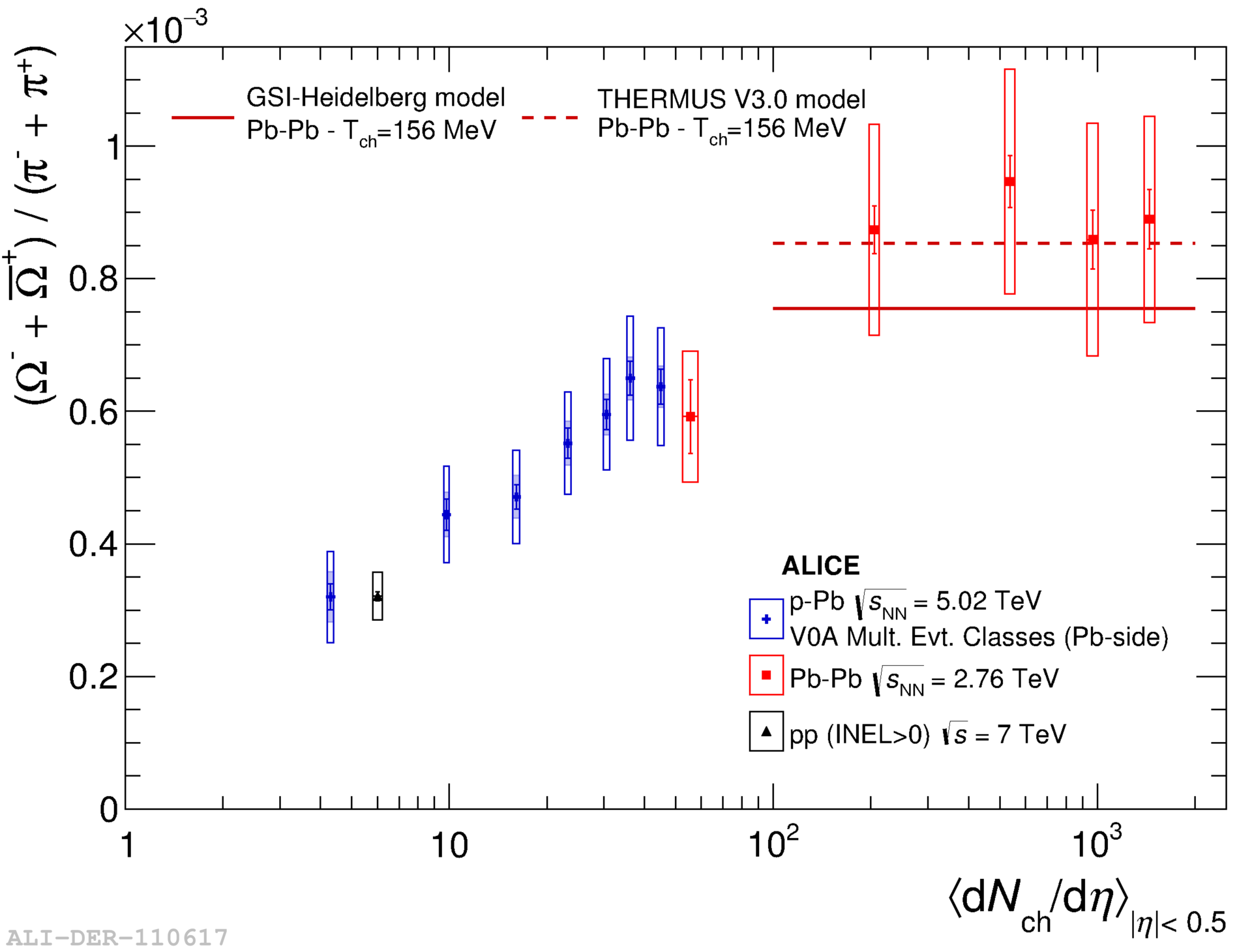}
\caption{\mbox{2$\Lambda$/($\pi^{-}$+$\pi^{+}$)} (top left), \mbox{($\Xi^{-}$+$\overline{\Xi}^{+}$)/($\pi^{-}$+$\pi^{+}$)} 
         (top right) and \mbox{($\Omega^{-}$+$\overline{\Omega}^{+}$)/($\pi^{-}$+$\pi^{+}$)}  (bottom) ratios as a function of 
         $\langle{dN_{ch}}$/d$\eta\rangle_{|\eta|<0.5}$ for \mbox{p--Pb} (filled blue cross) and \mbox{Pb--Pb} (filled 
         red square) collisions. Open boxes are total systematic uncertainties and shaded boxes show the 
         uncorrelated-over-multiplicity component. Lines represent predictions from statistical hadronization models 
         \cite{GSImodel} (continuous) and \cite{THERMUS} (dashed).}
\label{fig-3}
\end{figure}

The strangeness enhancement is known as one of the proposed signatures for the QGP formation in relativistic heavy-ion
collisions. Rafelski and M\"uller's expectation, first proposed in 1982 \cite{Rafelski}, is that QGP formation should lead to a 
higher abundance of strangeness per participating nucleon than in pp interactions. This phenomenon has been
actually observed at SPS \cite{strangEnhancNA57}, RHIC \cite{strangEnhancSTAR} and the LHC \cite{multsbPbPb}, and has been found to
increase with centrality and with the strangeness content of the particle, and decrease as the centre-of-mass energy increases.
Statistical hadronization models based on a grand-canonical approach have been demonstrated to be able to predict
particle yield ratios in heavy-ion collisions over a large energy scale \cite{thermalRHICvsLHC}. In this description, the energy 
dependence of strangeness enhancement has been understood as the consequence of a suppression of strangeness 
production due to the reduced phase-space volume in reference pp collisions (canonical suppression) \cite{cansup}.
At SPS and RHIC the strangeness enhancement has been studied looking at the ratio between the yield in \mbox{nucleus--nucleus} 
collisions and those in pp interactions at the same energy normalized to the mean number of participants 
($\langle{N_{part}}\rangle$). It has been shown that this is not the ideal way to isolate the enhancement component due to 
strangeness content, since the production rates of charged particles do not scale linearly with $N_{part}$ \cite{centrality}. 
A better observable is the ratio to pion yields, shown in Fig. \ref{fig-3} for $\Lambda$, $\Xi$ and $\Omega$ as a function of 
$\langle{dN_{ch}}$/d$\eta\rangle_{|\eta|<0.5}$ for \mbox{Pb--Pb}, \mbox{p--Pb} and pp collisions. In the case of 
multi-strange baryons, the ratio is shown to increase by up to a factor of about 2-3 going from pp collisions to 
\mbox{Pb--Pb} collisions and is more pronounced for the $\Omega$; for the $\Lambda$ a possible increase is not significant within
the systematic uncertainty. It is shown that the predictions from statistical hadronization models using a chemical 
freezeout temperature of 156 MeV \cite{GSImodel,THERMUS} are comparable with the ratios measured in the most central
\mbox{Pb--Pb} collisions for all the three particles. 
Looking at the \mbox{p–-Pb} data, an increase of the (multi-)strange baryon-to-meson ratios with multiplicity is observed.
The increase is seen to be more pronounced for particles with a larger strangeness quantum number.
In particular ($\Lambda+\overline{\Lambda}$)/$\pi^{\pm}$ and $\Xi^{\pm}$/$\pi^{\pm}$ ratios slightly exceed the 
saturation limit observed for \mbox{Pb--Pb}, while the $\Omega^{\pm}$/$\pi^{\pm}$ ratio is not higher than the one observed 
in peripheral \mbox{Pb--Pb}.
Comparison of hyperon-to-pion ratios as a function of pion multiplicity to the trends observed in 
statistical hadronisation models, where the local conservation of the strangeness is required (as in a canonical ensemble),
indicates that the behaviour is qualitatively consistent with the lifting of canonical suppression with increasing 
multiplicity \cite{multsbpPb}.

\section{Conclusions}
The production of (multi-)strange particles was measured in \mbox{Pb--Pb} and \mbox{p--Pb} collisions with the 
ALICE detector. 
The multiplicity dependence of the ($\Lambda$+$\overline{\Lambda}$)/K$^{0}_{S}$ ratio and the hardening of the spectra with 
increasing multiplicity point to similar collective behaviour. As observed for unidentified particles, $R_{\rm AA}$ for $\Xi$ is 
clearly suppressed at high $p_{\rm T}$, while $R_{\rm pPb}$ does not show any suppression.
The fact that relative strangeness production in \mbox{p--Pb} not only increases with multiplicity but also bridges pp and 
\mbox{Pb--Pb} collision values is an important recent observation in the study of strangeness production


\end{document}